# Current Flow Mapping in Conducting Ferroelectric Domain Walls using Scanning NV-Magnetometry

*Conor J. McCluskey*[*,1]*, James Dalzell*[1]*, Amit Kumar*[1]*, J. Marty Gregg**[*,1]

C. J. McCluskey: conor.mccluskey@qub.ac.uk

J. Dalzell: jdalzell06@qub.ac.uk

A. Kumar: a.kumar@qub.ac.uk

J. M. Gregg: m.gregg@qub.ac.uk

[1]Centre for Quantum Materials and Technologies, School of Mathematics and Physics, Queen's University Belfast, University Road, Belfast, BT7 1NN, United Kingdom.

## Abstract

The electrical conductivity of parallel plate capacitors, with ferroelectric lithium niobate as the dielectric layer, can be extensively and progressively modified by the controlled injection of conducting domain walls. Domain wall-based memristor devices hence result. Microstructures, developed as a result of partial switching, are complex and so simple models of equivalent circuits, based on the collective action of all conducting domain wall channels acting identically and in parallel, may not be appropriate. Here, we directly map the current density in ferroelectric domain wall memristors *in-situ,* by mapping Oersted fields, using nitrogen vacancy centre microscopy. Current density maps were found to directly correlate with the domain microstructure, revealing that a strikingly small fraction of the total domain wall network is responsible for the majority of the current flow. This insight forces a two order of magnitude correction to the carrier densities, previously inferred from standard scanning probe or macroscopic electrical characterisation.

## 1. Introduction

In ferroelectrics, domain walls are interfaces that separate volumes of differently oriented electrical polarisation (domains). If domain walls host a divergence of polarisation, such as when dipoles in neighbouring domains abut in "head-to-head" or "tail-to-tail" configurations, then an interfacial build-up of bound charge is expected. This bound charge had been theoretically predicted to drastically enhance local electrical transport, as early as in the 1970's[1]. However, while experimental observations of distinct electrical behaviour at domain walls were subsequently reported[2,3], it was not until 2009 that the concept of conducting domain walls became firmly established. Work by Siedel *et al*. showed, unequivocally, that regions of enhanced conductivity correlated fully with the locations of ferroelectric domain walls[4]. Since then, domain wall conductivity has been seen in a wide variety of ferroelectric systems[5–14]. Moreover, new device concepts have arisen, in which wall conductivity has been harnessed alongside the other key characteristic feature of ferroelectric domain walls: that they can be created, destroyed, or moved from place to place on demand. Progress has been quite rapid: demonstrator diodes[15], logic gates[16], memristors[17,18], rectifiers[19], transistors[20], memory devices[21] and neuromorphic elements[22] have all been reported, based on the active deployment, removal or modification of conducting domain wall pathways between source and sink electrodes.

Progress in uncovering fundamental physics of charge transport at conducting domain walls has been rather less rapid, for several reasons. Firstly, the mainstay tool in most of this kind of research (conducting atomic force microscopy, cAFM) is not suitable for the extraction of fundamental electronic transport properties, because of its two-probe nature: observed current maps depend on the tip-sample contact resistance which will vary considerably. Secondly, subsurface domain wall morphologies are unseen and often quite complex, meaning that the active current pathway is unknown and not easily controlled to create the standard geometries required for electrical resistivity or quantitative magnetotransport measurements.

Recently, there has been a small uptick in the number of studies seeking to get around these problems. They generally attempt to measure potential drops associated with the driving electric field along a current carrying domain wall, or, in the presence of a magnetic field, the developed Hall potential. In complex domain wall networks, these potentials have been measured spatially using Kelvin probe force microscopy (KPFM)[11,23–25], and then correlated with the relevant domain structure, whereas in the case of well-defined domain wall morphologies, relevant potentials can be measured using conventional evaporated point contacts[26–28]. While limited in number, these studies have allowed for crucial information such as charge carrier sign, density and mobility to be established, though they often rely on assumptions of current carrying pathways.

In lithium niobate (LNO) single crystal thin films, the two-probe conductance of top-down parallel plate capacitor structures can be modified by up to twelve orders of magnitude[18] by sequentially increasing the number and inclination angle of conducting domain walls. This extraordinarily tuneable change in conductance results from a dense network of conducting domain walls, which straddles the interelectrode gap of the capacitor, combined with distinct inherent transport property differences between the domain walls and the domains themselves (which are extremely strongly insulating). Domain wall transport measurements in this system[28,29] suggest room temperature electron carrier mobilities between 300 and 3700 $cm^2V^{-1}s^{-1}$, among the highest recorded in oxide interfaces to date. Such device tunability, combined with exciting transport properties of the domain walls, has placed LNO at the forefront of domain wall device research[16,18,30,31]. The complexity of the network of conducting domain walls in this system, however, makes quantitative deductions about the active conducting area and estimates of associated basic transport properties (such as conductivity and carrier density) of domain walls difficult.

Herein, we report the use of a single nitrogen-vacancy defect spin state, within the diamond tip in a nitrogen-vacancy scanning probe microscope (NV-microscope), to determine the Oersted field associated with a current carrying domain wall network in an LNO-based parallel-plate capacitor / memristor. By inverting the Biot-Savart law, quantitative maps of the two-dimensional current density vectors, which are the source of the measured Oersted fields, have been generated[32–34]. Analysis of such current density maps, considered alongside known domain wall microstructures, have revealed that a surprisingly small fraction of domain walls are strongly active in conduction. With this information, we have revisited domain wall conductivity and carrier density estimates, and discussed new values, in the context of all available transport information in conducting domain wall systems.

## 2. Results

### 2.1 Domain wall injection

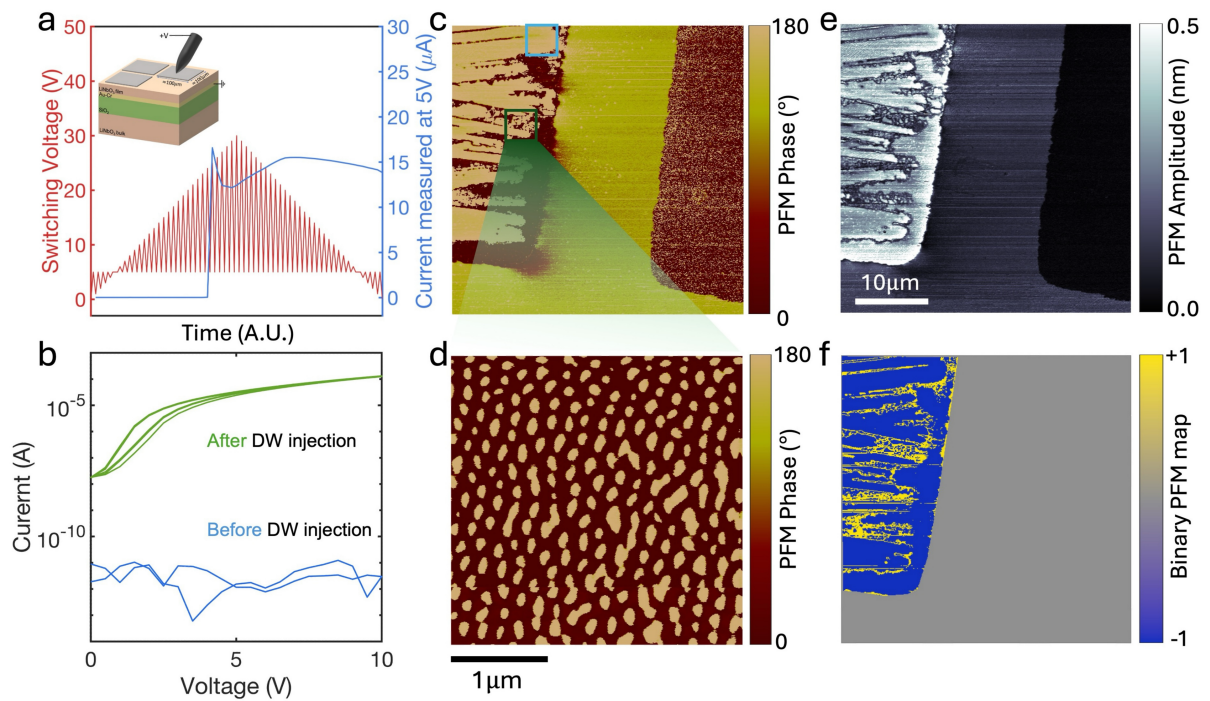

**Figure 1: Domain wall injection and microstructural investigation. a,** The switching voltage pulse and corresponding read current (measured at 5V) for the domain wall injection procedure. The inset shows a schematic of the LNO capacitor structures. **b,** IV curves taken before and after domain wall injection, at room temperature. **c,** PFM phase image of the top surface of the LNO capacitor structures, taken after partial reversal of polarisation by application of the switching pulse, presented in **a**, to one of the sputtered square electrodes (left hand side). **d,** PFM phase map, at a greater magnification, showing the complex, circular domain structure typically found after the partial poling of LiNbO$_3$ films. **e,** PFM amplitude map corresponding to **c**. **f,** A binary domain wall map generated by masking the electrode area and selecting domain wall pixels from the amplitude map in **e**, as described in the main text.

LNO is a uniaxial ferroelectric with only two possible polarisation orientations, lying along the crystallographic *c*-axis. It is an exceptional electrical insulator, with a room temperature bulk electrical conductivity[35,36] of less than $10^{-17} \Omega \text{cm}^{-1}$. This can be drastically altered by the presence of charged, conducting 180 degree domain walls, which provide pathways of enhanced electrical conduction through the insulating bulk matrix[6,18,37,38]. **Figure 1** illustrates the point. The inset to panel **a** shows a schematic illustration of LNO capacitor structures, generated by sputtering of planar, platinum thin film electrodes onto the surface of commercially obtained (from NanoLN) z-cut ion sliced lithium niobate films (~500nm in thickness). The as-received LNO films have a gold-chromium bottom electrode, and are initially monodomain, with polarization pointing out of the plane of the film. Panel **a** also shows the triangular voltage pulse applied to the top electrodes of a typical LNO capacitor structure, to partially reverse the polarization, and inject conducting domain walls into the capacitor structure, as discussed extensively in previous work[16,18,22,29]. After each increasing voltage pulse ("switching pulse"), a 5V "reading pulse" is applied. The measured current ("read current") is an indication of the conductance state of the capacitor (or memristor). It remains at the noise level, until the switching pulse reaches approximately 26V (an equivalent electric field of ~5 x $10^7$ Vm$^{-1}$), after which it increases by several orders of magnitude. Such an increase in current is indicative of conducting domain wall pathways within the ferroelectric. In panel **b**, IV curves before (blue) and after (green) the switching procedure show a persistent, steady-state increase in conductance of approximately 8 orders of magnitude (as measured at 10V). Panel **c** shows a piezoresponse force microscopy phase image, taken while scanning over the platinum top electrode of one of the parallel plate capacitor structures which has undergone the switching procedure in **a**. A rich network of domains and domain walls exists underneath the poled electrode. Some regions (for example, that enclosed within the blue box in **c**) appear to be fully inverted monodomain regions, spanning mesoscopic scales. Other regions consist of fine mixed domain states with dense networks of circular domain walls. A typical mixed domain state is shown with higher resolution PFM in panel **d**. Such a microstructure is commonly observed in poled thin-film LNO[16,18,22,29]. Studies of the subsurface domain wall morphology, in both bulk[6,39] and thin film[18,29] geometries, agree that conducting domain walls formed by polarisation reversal are usually inclined with respect to the polar axis, though to varying degrees: in the thin films used here, walls have inclination angles on the order of 10-20 degrees (see the cross sectional TEM images in **figure S1c**), while the inclination angle in bulk single crystals is smaller (typically between 1-5 degrees[37]).

Inclination angle, which determines the magnitude of the polarisation divergence at the wall, determines conductivity: higher inclination results in higher conductance[28,37]. Even in the apparent monodomain regions, the subsurface domain network is non-trivial (in the case of domains with inverted polarisation). The mesoscopic domains, observed on the surface, are actually a result of the coalescence of many domains with fine circular microstructure, remnants of which can linger just under the surface of the film. This can be seen in the tomographic PFM in figure **S1a-b**, where an apparent single domain region breaks up very quickly into a fine circular microstructure, after removing approximately the top 50nm of the film surface by AFM machining. This microstructural complexity makes estimating the current carrying area difficult, though an initial estimate can be made from the fractional coverage of domain walls underneath the electrode.

Pixels corresponding to domain walls have been selected from the PFM amplitude map in figure **1e** by first masking the electrode area and then selecting pixels with values which are lower than a cutoff amplitude value. The result is the binary domain wall map shown in panel **f:** pixels with values of +1 (yellow) correspond to the pixels that contain a domain wall; values of -1 (blue) correspond to pixels within +*P* or -*P* domains, and values of 0 (grey) correspond to areas outside the partially switched capacitor-memristor. We can readily calculate the fractional coverage of pixels containing domain walls, $f_{DW}$, under the electrode:

$$f_{DW} = \frac{n(+1)}{n(+1) + n(-1)} \approx 0.24 \qquad (1)$$

This can be used to estimate the total area of the capacitor-memristor corresponding to the intersection of the domain walls with the electrode-ferroelectric interface, $A_{DW}$:

$$A_{DW} = f_{DW} * A_E * \frac{t_{DW}}{t_{Pix}} \approx 3.8 \times 10^{-11} \, m^2 \qquad (2)$$

Where $A_E$ is the area of the electrodes (110 μm x 110 μm). The final term in equation (2) $\left(\frac{t_{DW}}{t_{Pix}}\right)$ accounts for the fact that domain walls are likely to be much thinner than a pixel, in a typical PFM scan, and so the domain wall area is smaller than the total area of the pixels in which domain walls are detected. Here, the pixel "length" ($t_{Pix}$) is 78nm, and the domain wall thickness ($t_{DW}$) is assumed to be 1nm. We next investigate the current carrying characteristics of the domain walls.

## 2.2 *In-situ* Characterisation of Domain Wall Capacitors

*AFM-based Electrical Studies*

The measured current can be spatially correlated with the domain walls by means of cAFM, shown in **Figure 2**. Panel **a** shows a schematic of the measurement. The sputtered top electrode of a partially poled capacitor structure was removed by AFM micromachining. Matching PFM domain maps (**b**) and conductive AFM maps (**c**) shows two things: firstly, as expected, the regions of enhanced conductivity correspond to the locations of domain walls; secondly, the vast majority of domain walls visible in PFM also appear as regions of enhanced conductivity in conductive AFM. This suggests that most of the domain walls visible in surface level PFM electrically contact the bottom electrode, and, therefore, constitute fully penetrating conductive conduits with the capability to carry current. So, in the capacitor geometry, all of the domain walls should act in parallel, contributing approximately equally to the total current driven through the film, provided all domain walls experience the same magnitude of electric field.

To check this, we next investigated the current carrying characteristics of the capacitor structures *in-situ*, using Kelvin probe force microscopy (KPFM). In panel **d**, we show a schematic of the measurement. Domain walls were injected into one parallel plate capacitor by the procedure outlined in the previous section. Then, this top electrode was electrically connected to several others via focussed ion beam deposition of conducting platinum interconnects. This allowed an electrical bias to be supplied to the relevant top electrode, without obscuring the electrode surface from the scanning probe microscope. Since none of the other series-connected electrodes contacted subsurface conducting domain walls, and bulk LNO is exceptionally insulating, leakage current from the connected row of electrodes to the bottom electrode is negligible. In other words, the series of electrodes simply acts as electrical channel to the final, "active" electrode, which contains the domain wall pathways needed to allow percolation of current through the circuit. Topography (**e**) and potential maps (**f**) encompassing the final interconnect and the top electrode of the active capacitor show that the electrode area is equipotential, approximately matching the externally applied bias of 1V. This suggests that the domain walls underneath the entire electrode are indeed subject to a uniform electrical field. Note that while uniform, the driving electric field can be quite reduced, due to a significant contact resistance between the top electrode and the domain walls. Indeed, the non-linearity of the 2-probe IV curve in figure 1 suggests a contact resistance is present (if the wall behaviour is assumed to be Ohmic). Four-probe measurements are challenging in top-down domain wall geometries, so as a first approximation, we proceed with the estimation that the driving electric field is $E = 200$ kVcm$^{-1}$ (given simply by the applied voltage, 10V, divided by the film thickness, 500nm). With the microstructural investigations of figure 1, and the electrical characterisations in figure 2, we can make an estimate of the implied carrier density and conductivity of the domain wall system. Given the current driven across the domain wall network in figure **1a** ($I \approx$132µA at 10V), the 2D carrier density is

$$n_{2D} = n_{3D} * t_{DW} = \frac{\sigma}{e\mu_e} t_{DW} = \frac{I}{A_{DW} e \mu_e E} t_{DW} \approx 3 \times 10^5 \text{ cm}^{-2} \quad (3)$$

Where $e$ is the electronic charge and $\mu_e$ the electronic mobility. Here, we use a value of $\mu_e = 3700$ cm$^2$V$^{-1}$s$^{-1}$ as implied by geometric magnetoresistance measurements in similar samples[29]. Hall effect measurements in x-cut ion sliced LNO[28] suggest a slightly lower value of $\mu_e \approx 337$ cm$^2$V$^{-1}$s$^{-1}$; taking this value increases the carrier density estimate by one order of magnitude. Note that a 2D carrier density is independent of the assumed domain wall thickness ($t_{DW}$ also appears in the expression for $A_{DW}$). The implied carrier density is extremely low, much smaller than that typical for metallic 2DEGs, and many orders of magnitude smaller than that expected from full screening of the polar discontinuity:

$$n_{screen} = \frac{2\boldsymbol{P} \cdot \boldsymbol{S}}{eS} = 1.8 \times 10^{14} \text{ cm}^{-2} \qquad (4)$$

Equation (3) probably underestimates the carrier density, as it neglects the contribution of a contact resistance between the electrodes (the genuine driving field is probably lower than 200 kVcm$^{-1}$). Nonetheless, correcting for contact resistance is unlikely to account for the 9 order of magnitude difference between the assumed screening charge and the measured carrier densities.

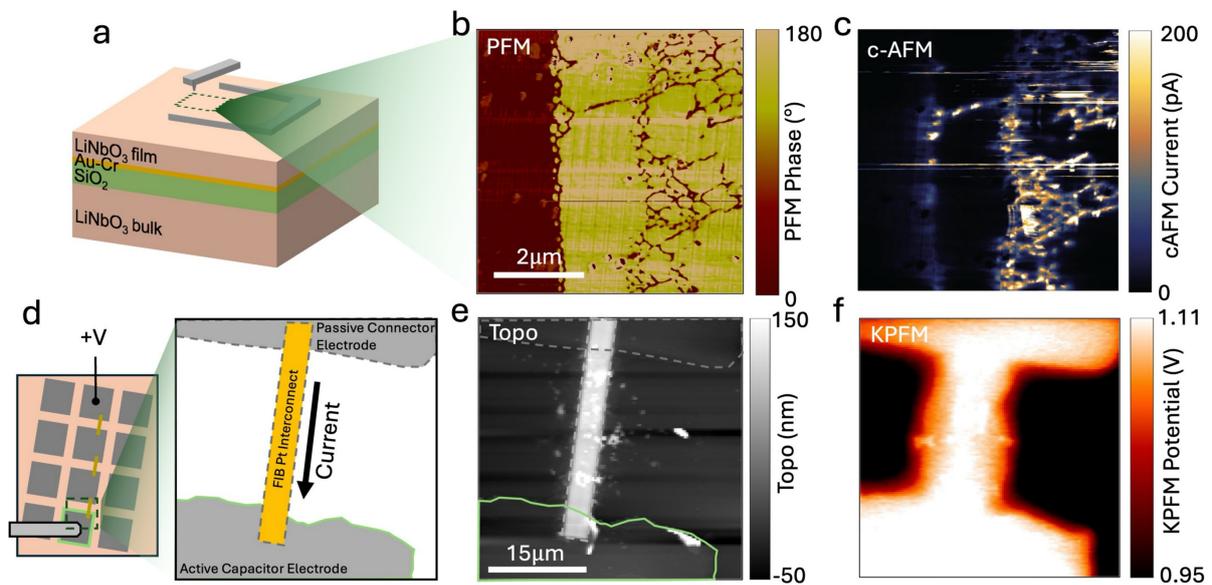

**Figure 2: Scanning probe investigations. a,** A schematic showing the AFM machining technique used to allow scanning probe access to the domain microstructure under the top electrode of an LNO capacitor structure. **b,** PFM phase and **c,** corresponding conductive-AFM maps of the revealed domain microstructure. **d,** A schematic of the series of connected top electrodes, linked by FIB deposited Pt interconnects, which was used to investigate the surface potential of the active capacitor *in-situ*. The active capacitor top electrode has a solid green outline. **e,** Topography and **f,** measured surface potential, taken during a Kelvin probe force microscopy scan, with 1V applied to the series of connected top electrodes.

*In-situ Current Density Mapping using NV-Centre Microscopy*

Two dimensional current density distributions can be uniquely reconstructed from measurements of a single component of the generated Oersted field, by inversion of the Biot-Savart law (described in detail in supplementary section S2)[33,34] . Such measurements can be made by scanning nitrogen vacancy (NV$^-$) centre magnetometry: a single NV$^-$ centre defect is implanted into a diamond scanning probe tip, and is raster-scanned over the surface of the sample, measuring the projection of magnetic field along the NV-axis at each point. Figure **3a** shows a schematic of scanning-NV measurements performed on one of our current-carrying LNO capacitor structures *in-situ*. Conducting domain wall pathways exist over the entire electrode area (shown in supplementary figure **S2a-b**). Panel **b** shows the measured magnetic field map close to the current carrying interconnect, which links the final inactive top electrode to the top electrode of the active capacitor. The magnetic field signal is strongest surrounding the Pt interconnect, and decays very quickly as current enters the larger capacitor top electrode. No notable features are seen in a scan encompassing a much larger region of the electrode area than that shown in figure **3b** (see figure **S2** of the supplementary), so instead we focus on the region very close to the current carrying interconnect. The Oersted field is related to the current density, so this drop in field information is expected as current spreads out upon entering the wider electrode. The logarithm of the magnitude of the reconstructed 2D current density is shown in the colourmap of panel **3c,** along with the in-plane vector components represented as an overlayed quiver plot. A corresponding drop in current density strength is observed. Interestingly, there is some clear structure to the current density pathways as current enters the top electrode in the active memristor device. This is seen most clearly in the current density contour map, shown in panel **3d.** Three distinct current channels are observed, represented by the three overlaid green arrows. The approximate location of the platinum inlet is illustrated by the dotted black line. Panel **3e** shows an overlay of these current pathways with a PFM amplitude map taken from the same location, after removal of the top electrode by AFM machining. There is a clear correspondence between the current density channels and the local domain microstructure: current channels towards the highly dense regions of domain wall texture in the immediate vicinity of the Pt interconnect. Note that in panel **c,** the two-dimensional current density has been divided by the approximate thickness of the electrode (100nm) prior to taking the logarithm, giving units of A/cm$^2$. This is equivalent to the assumption that the current density is uniformly distributed along the z-direction.

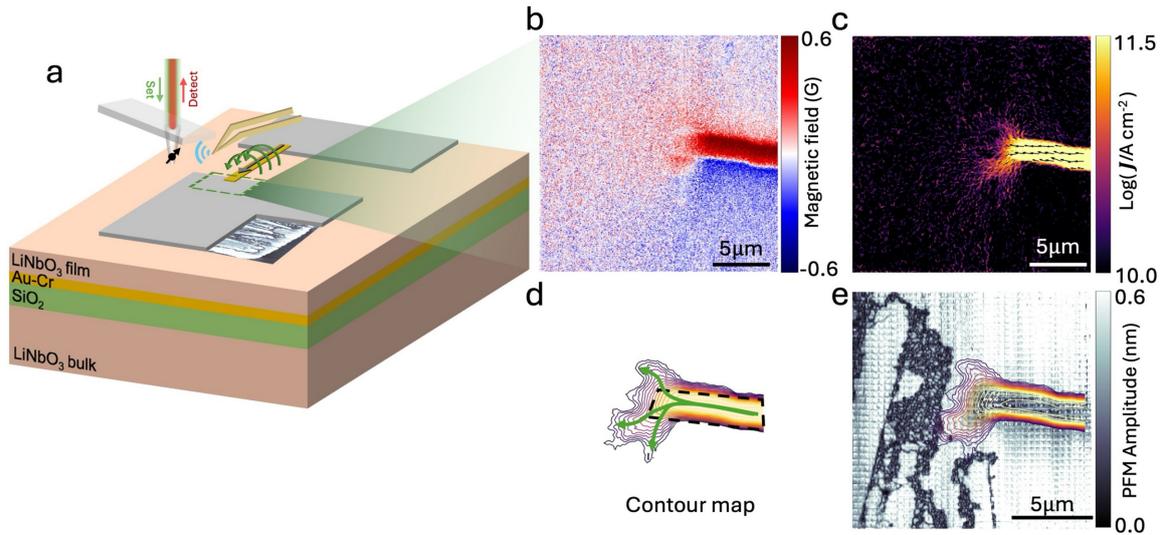

**Figure 3:** *In-situ* magnetic field mapping using NV-centre microscopy. **a,** A schematic showing the *in-situ* NV-centre magnetometry mapping. Current is supplied through the series connected top electrodes and enters the capacitor top electrode via the Pt interconnect (grey arrow). Green arrows represent the Oersted field associated with the current. **b,** Measured NV-centre magnetic field map, taken in the vicinity of the Pt interconnect (green dotted box in **a**). **c,** Logarithm of the reconstructed 2D current density map, generated from the measured magnetic field map in **b**. **d,** A contour map generated from the current density amplitude, showing the channelling of current. **e,** An overlay of the local ferroelectric domain structure (PFM amplitude) with the current density contour map.

## 2.4 Finite Element Modelling

To understand the implications of the observations made, we compare our data to results from finite element modelling of current carrying structures, distributed under a top electrode (using COMSOL Multiphysics). We model a metallic, current carrying track which contacts a larger, planar top electrode. The top electrode is separated from a larger, planar bottom electrode by a 500nm thick electrically insulating block, which plays the role of bulk LNO. Embedded within the insulating block are several conducting channels, mimicking the domain walls. These are modelled simply as metallic cylinders, contacting the top and bottom electrode; their conductivity is smaller than that of the top surface track and electrodes by 4 orders of magnitude (based on previous estimates of domain wall conductivity in LNO thin films[29,38]). In the first instance, illustrated in figure **4a**, we insert only three conducting channels distributed close to the current-carrying inlet. Since these three channels constitute the only pathways for current flow between the top and bottom electrode, the modelled current density clearly reflects the geometry of the conducting channels: upon entering the top electrode, the total current density splits into three distinct pathways, each oriented towards one of the conducting channels. This is clear from both the colour intensity map of current density magnitude, and corresponding contour map (panels **4b** and **4c** respectively)**.**

Arrays of additional conducting channels, with the same conductivity, were then added, distributed throughout the entire capacitor area (as illustrated in figure **4d**). The corresponding modelled current density profile (figures **4e,f**) is qualitatively rather different, showing a much more homogeneous spread of current as it leaves the inlet to the top electrode, due to collective action of all domain walls acting perfectly in parallel. The experimentally measured current density contour map, replotted in figure **4d**, displays obvious evidence of channelling towards the nearest available domain walls, and so is indicative of the first modelled scenario, as opposed to the second. This is despite the fact that the ferroelectric microstructure in the thin film more closely reflects that of the second model, where conductive channels are distributed throughout the entire electrode area (figure **2**). The experimentally determined information, taken alongside the modelling expectations, indicates that only a small subset of walls near to the current inlet to the top electrode are strongly active in conduction. If the conductivity of the conducting channels is made equal to or higher than that of the metallic contacts, then the characteristic features of the maps in figure **4b,c** are recovered (see supplementary figure **S3b**). However, this scenario is unlikely for two reasons. Firstly, the currents measured indicate resistances on the order of 100kΩ, which is inconsistent with domain wall conductivities (and associated contact resistances) of the same magnitude as the metallic contacts. Secondly, the potential profile seen on the top electrode would no longer be equipotential in this case (see figure **S3c-d**): this directly contradicts the measured potential profiles in figure **2f.** Nonetheless, the measured NV current density profiling in figure **3** clearly suggests that only a local subset of domain walls, visible in figure **3d**, is active or dominant in current conduction. In figure **4h**, the domain wall area has been recalculated, this time considering only this subset of domain walls. Once again, the number fraction of pixels containing domain walls is calculated as:

$$f'_{DW} = \frac{n'(+1)}{n'(+1) + n'(-1)} \approx 0.22 \tag{5}$$

and, following from equation (2):

$$A'_{DW} \approx 5.8 \times 10^{-13} \ m^2 \tag{6}$$

giving a corrected 2D carrier density:

$$n'_{2D} \approx 1.9 \times 10^7 \ cm^{-2} \tag{7}$$

which is two orders higher than that inferred from the domain mapping and IV response in figures **1** and **2**.

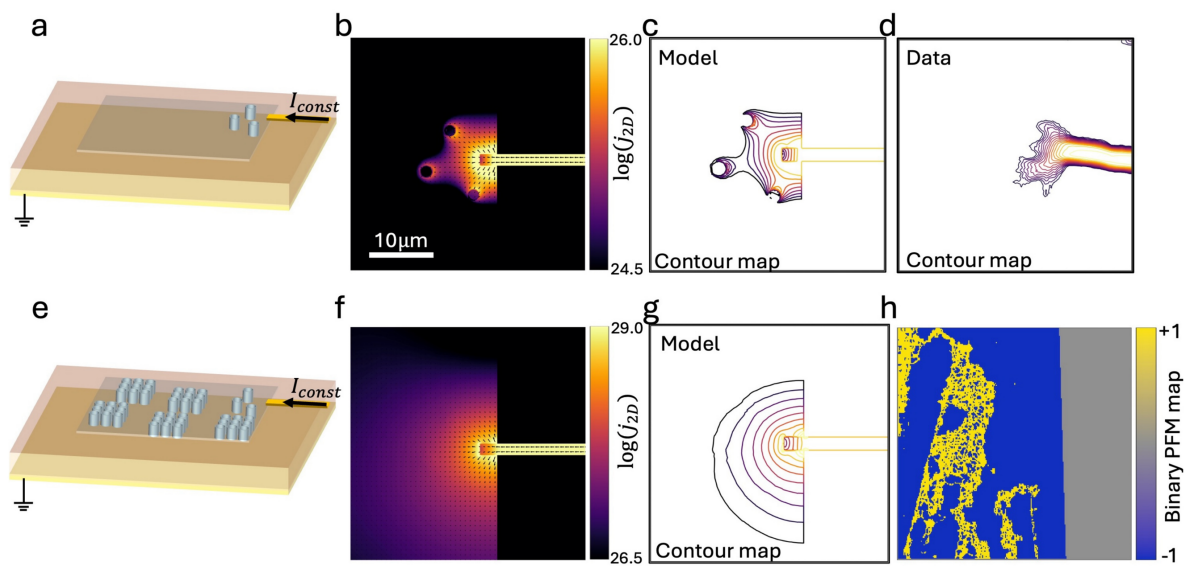

**Figure 4: Finite element modelling. a,** Schematic of the finite element model where only a small number of percolating pathways exists between the plates of a capacitor. **b,** Current density extracted from the modelling **c,** Contour map extracted from the modelled current density. **d,** Current density contour map extracted from the measured data in figure 3. **e, f, g,** show the same information as **a,b,c**, but when many domain wall pathways are active, as illustrated in **e. h,** The binary domain wall map corresponding to the subset of domain walls active in conduction, as suggested by the current channelling seen experimentally.

## 3. Discussion

In table 1, we have collated parameters from all transport measurements made in conducting domain wall systems, across a number of publications. Where needed, the active carrier densities are converted to 2D estimates by multiplying by the authors' assumed DW thickness, giving the two-dimensional active carrier density, $n_{2D}$. Screening carrier densities ($n_{screen}$) are calculating using equation (4), with values for the spontaneous polarization (at room temperature) of 70 µC cm$^{-2}$, 26 µC cm$^{-2}$ and 6 µC cm$^{-2}$ for LiNbO$_3$ , BaTiO$_3$ and ErMnO$_3$, respectively[26,40,41]. Uniaxial ferroelectrics are most commonly employed for domain wall transport studies, presumably due to the relative simplicity of a domain configuration containing only two polar variants. The magnitude of the polar discontinuity at the domain wall is then solely determined by the domain wall inclination angle, $\alpha$, which is defined here as the angle subtended between the wall surface normal and the polarisation axis. A domain wall inclination angle of $\alpha = 0°$ is associated with the maximum magnitude in polarisation divergence, and $\alpha = 90°$ indicates no polarisation divergence at the wall. The only exception is tetragonal $BaTiO_3$, where the abutting polarisations in neighbouring domains are at 90° to each other.

| System | $n_{2D}$ (cm⁻²) | $n_{screen}$ (cm⁻²) | Carrier mobility (cm²V⁻¹s⁻¹) | Comment |
|---|---|---|---|---|
| 180° DWs, z-LiNbO₃ 500nm film (this work) | $2 \times 10^7$ | $2 \times 10^{14}$ | 3700 | Geometric magnetoresistance measurements in ion-sliced thin film LNO. DW inclination angle ≈78° [from [29]]. |
| 180° DWs, x-LiNbO₃ 500nm film (5% MgO) | $3 \times 10^6$ | $9 \times 10^{14}$ | 337 | Hall effect measurements in ion-sliced thin film LNO. DW inclination angle 0° [from [28]]. |
| 180° DWs, z-LiNbO₃ Single crystal (5% MgO) | $2 \times 10^4 - 3 \times 10^5$ | $8 \times 10^{13}$ | 35-54 | Hall effect measurements on congruent single crystal LNO. DW Inclination angle >86° [from [27]]. |
| 90° DWs, BaTiO₃ Single Crystal | $7 \times 10^3$ | $3 \times 10^{14}$ | 395 | Mobility measured by the Hall effect [from [26]]. |
| 180° DWs, ErMnO₃ Single Crystal | $1 \times 10^7$ | $7 \times 10^{13}$ | 670* | Mobility measured by AFM-based Hall effect. Meandering DW structure, inclination angle 0° at most charged points [from [23]]. |

*Table 1*: A collection of charge mobility and carrier density estimates across conducting domain wall literature. *Hole mobility.

The most striking trend in table 1 is the fact that measured carrier densities are consistently orders of magnitude lower than the required screening carrier density. The relatively broad spectra of systems, doping levels and domain wall inclination angles suggest that this is a general feature of domain wall transport. The carrier mobilities, all measured at room temperature, are also generally quite high. The lowest values are seen in bulk LNO, which has low bound charge, owing to the shallow domain wall inclination. It has been demonstrated explicitly, in several instances, that domain wall conductivity (or, conductance) is dependent on inclination angle. Typically, the assumed rationale is that more inclined domain walls host larger polar discontinuities, and that this should result in a higher active carrier density present at the domain wall. Qian *et. al* demonstrate[28] an explicit $\sigma \propto \sin \theta$ relationship for H-H domain walls in *x*-cut ion-sliced lithium niobate domain walls with controlled inclination angle, which suggests a conductance scaling with bound charge density, while Beccard *et. al* measure differing carrier densities for two samples with different distributions of inclination angle[27]. Nonetheless, looking at $LiNbO_3$ specifically, the variation of measured carrier mobility in table 1 is as significant as the variation in active carrier density; this begs the question as to whether the intrinsic properties of the carriers (such as effective mass / mobility) are affected by domain wall inclination angle, and that the assumption of carrier density alterations might be overly simplistic. In other words, if such a small fraction of screening carriers (one in every $10^7$) contribute to conduction, then a change in domain wall inclination angle, which affects the screening carrier density, might not affect the *active* carrier density as much as has been assumed to date. Of course, the above data is generated from different samples with different doping, and presumably different defect levels, which can all contribute to scattering and hence mobility differences, making direct quantitative comparisons difficult.

Finally, we note that the active carrier densities suggested are extremely low, in comparison to those generally observed in 2D systems such as metallic 2DEGs or 2D materials (less than $10^{12} cm^{-2}$ is considered reasonably low). Usually, such low carrier densities are desirable for studying unusual electronic correlation and localisation phenomena (for example Wigner crystallisation and metal insulator transitions)[42]. The high mobility and low carrier density transport, observed throughout domain wall research, is therefore perhaps more indicative of semiconducting behaviour.

## 4. Conclusions

We have studied the current carrying properties of conducting domain walls in insulating, ferroelectric lithium niobate. We have demonstrated direct imaging of the current density in LNO domain wall capacitor / memristor structures by NV-magnetometry, which reveals that the current flows selectively through a subset of domain walls present in the structure. This fact is missed by analysis with standard scanning probe techniques such as complementary PFM and c-AFM. Using these insights, we deduce that the active carrier density is significantly larger than prior estimates (by a factor of ~100). Finally, we have collated the available transport information from conducting domain wall systems and discussed several apparent features of domain wall transport: carrier mobilities are high, carrier densities are very low, and both properties seem to vary with inclination angle.

## 5. Experimental section

*Capacitor fabrication and poling:*
Square Pt electrodes of approximately 110µm side length were sputtered onto commercially available 500nm thick ion sliced z-cut lithium niobate, patterned through a copper TEM grid hard-mask. The as-received sample includes a Au-Cr bottom electrode. Isolated top electrodes were contacted by a tungsten probe, using a micromanipulator, and the voltage profile in figure 1 was supplied (and corresponding currents measured) using a Keysight B2910BL source measure unit. To ensure good electrical contact, the tungsten probe was coated in liquid metal gallium-indium-tin eutectic. A typical pulse length is between 0.1-1 second. A compliance current limit of around 1mA was set, which occasionally caused the voltage profile to depart from that represented in figure 1. The films are initially monodomain, with polarisation pointing away from the bottom electrode. The bottom electrode was grounded, and the supplied voltage positive, meaning the sense of the applied field is in opposition to the original polarisation.

*Atomic force microscopy methods:*

All atomic force microscopy data was taken using an MFP-infinity system (Asylum Research, Oxford instruments).

**Piezoresponse force microscopy**

Pt-Ir coated Si conducting AFM tips with a nominal free resonant frequency of 70kHz were used (Nanosensors, PPP-EFM). In figure 1, the PFM information was gathered through the top electrodes by application of the alternating voltage directly to the square electrode (using a series connection like that shown in figure 3D), with the tip acting as piezoresponse sensor only. The AC bias had an amplitude of 2V and a frequency tuned to the vertical tip-sample resonance, around 300kHz. The PFM in figure 3b was taken after removal of the top electrode, so the AC bias was supplied to the tip directly.

**Conductive AFM**

Conductive AFM imaging was performed using the same Pt-Ir coated conducting tips, scanned in contact mode. A DC bias of -7V was applied to the bottom electrode.

**Kelvin probe force microscopy**

KPFM imaging was performed using the same Pt-Ir coated conducting tips. Potential measurements were made using a dual pass mode, with a "delta height" of 50nm during the potential measurement.

*Scanning NV magnetometry:*

NV magnetometry measurements were made using a commercially available scanning NV magnetometer (ProteusQ from Qnami AG). The microscope enables optical readout of the spin state of a single NV-defect, embedded in a diamond pillar, while raster scanning over the surface of a sample using tuning fork-based AFM feedback. Here, we used Quantilever MX tips, addressed by continuous wave green laser light and microwave excitation. The magnetic field maps were taken in 'full-B' mode, meaning full electron spin resonance spectra were taken at each pixel, and the locations of dips in the fluorescence of the NV defect (in frequency space) are used to quantitatively determine the magnetic field along the NV-axis in the vicinity of the defect at each point. A current of approximately 400uA was supplied to the capacitor structure during NV scanning using a Keysight B2910BL source measure unit by application of a constant voltage, maintained below the coercive voltage of the films. The current reconstruction procedure (outlined in the supplementary info) was implemented in MATLAB.

*Finite Element Modelling:*

Finite element modelling was performed using commercially available modelling software (Comsol Multiphysics). The electric currents physics node of the AC/DC module was used to apply a constant current through edge of the top electrode, and an edge of the bottom electrode was held at ground.

## 6. Acknowledgements

J. M. G and C. J. McC are grateful for funding support from the Engineering and Physical Sciences Research Council (EPSRC), grant number EP/X027074/1 (CAMIE). A. K. gratefully acknowledges support from the Department of Education and Learning NI through grant USI-211. J. D. acknowledges support via the Engineering and Physical Sciences Research Council (EPSRC) under grant number EP/S023321/1. The authors would also like to acknowledge Dr. William Hendren and Dr. Jade Scott for assisting with thin film electrode deposition.

## 7. Conflict of Interest

The authors declare no conflict of interest.

## 8. Author Contributions

The capacitor preparation, poling, and standard characterisation were carried out by C. J. McC. NV-centre microscopy mapping and current reconstruction analyses were performed by C. J. McC and J.D.. C .J. McC. performed the finite element modelling. J.M.G. and A. K. contributed to the experimental design and interpretation of results. The experimental idea was conceived by J.M.G., who also supervised the research project. All authors contributed to the preparation of the manuscript.

## 9. Data Availability Statement

The data that supports the findings of this study are available at the following URL: XXXXXXXX, and at reasonable request from the authors.

# Supplementary Material: Current Flow Mapping in Conducting Ferroelectric Domain Walls using Scanning NV-Magnetometry

Conor J. McCluskey*,[1], James Dalzell[1], Amit Kumar[1], J. Marty Gregg[#,1]

Email: *conor.mccluskey@qub.ac.uk and #m.gregg@qub.ac.uk

[1] Centre for Quantum Materials and Technologies, School of Mathematics and Physics, Queen's University Belfast, University Road, Belfast, BT7 1NN, United Kingdom.

## Section S1: Additional Microstructural Information

As mentioned in the main text, the subsurface domain wall morphology in LiNbO$_3$ capacitors can be much more complex than surface imaging suggests. Figure **S1** shows PFM phase domain images, taken on the top surface of a switched LNO capacitor (**a**), and after removal of some material by AFM-micromachining (**b**). The green square in **a** and **b** mark the region which has been milled, which is a square of length $3\ \mu m$. At the top surface, the switched region appears as one relatively large domain, however machining reveals a that the structure actually consists of a series of conical domains and domain walls, which coalesce near the top surface. The cross-sectional TEM image shown in (**c**) shows a complementary view of the domain wall cones coalescing just below the film surface.

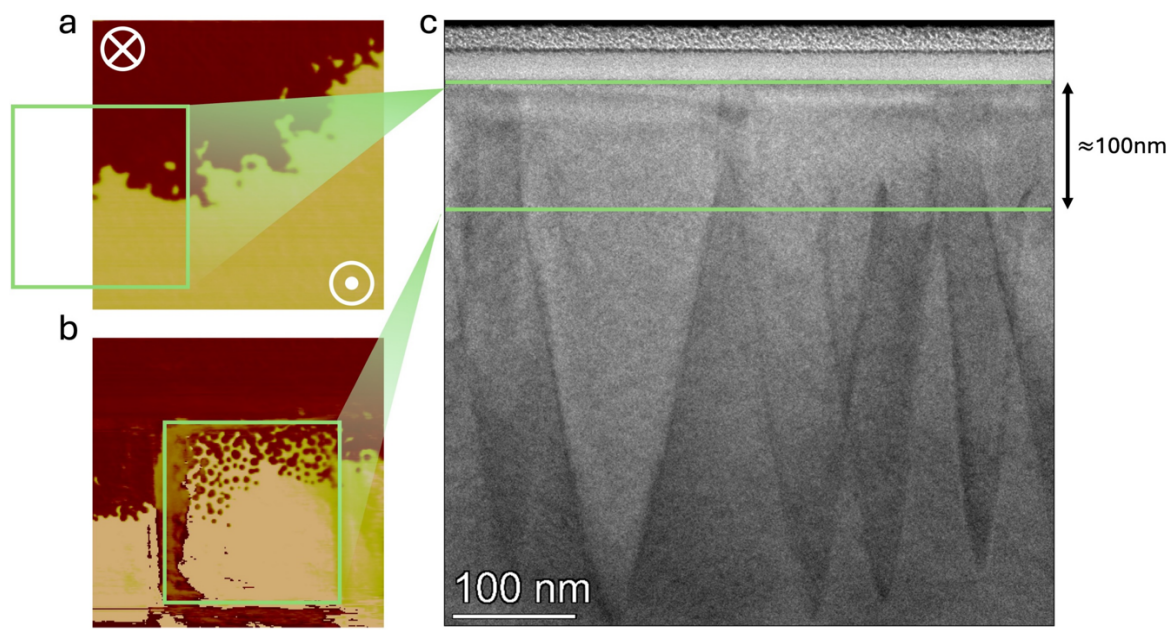

**Figure S1: Subsurface Microstructural Detail. a,** Vertical PFM phase information taken at the top surface of an LNO film which has been subjected to the switching procedure. The yellow regions experienced no electric field as they were outside the electrode area. The red contrast marks the switched region with reversed polarisation. **b,** PFM phase information taken after removing approximately 100nm of the LNO thin film by AFM machining. The green square is 3 $\mu m$ x 3 $\mu m$ and marks the machined region. The corresponding region is also marked in **a**. **c,** High-resolution, cross-sectional TEM imaging of a lamella taken from a poled area of the LNO film. The top green line marks the top of the film, and the second green line marks 100nm along the z-direction, approximately matching the depth of the AFM-machined trench in **b**.

## Section S2: Current reconstruction procedure

The discussion in this section follows refs [1–3]. The relationship between current density and the generated Oersted field is given by the Biot-Savart law:

$$\boldsymbol{B}(\boldsymbol{r}) = \frac{\mu_o}{4\pi} \iiint \frac{\boldsymbol{J}(\boldsymbol{r}') \times (\boldsymbol{r} - \boldsymbol{r}')}{|\boldsymbol{r} - \boldsymbol{r}'|^3} d^3 \boldsymbol{r}' \quad \text{(S1)}$$

Where $J(r')$ is the current density at the source location $r'$, $r$ location of magnetic field measurement, and $\mu_o$ is the permeability of free space. The expressions for the Cartesian components of the magnetic field are convolutions of the current density components, $J_x$, $J_y$ and $J_z$ with a Green's function that depends on the distance between the source and the magnetic field measurement $(\boldsymbol{r} - \boldsymbol{r}')$[1,2]. The cross product implies that $B_y$ contains information regarding $J_x$, and vice versa.

To reverse the convolution (and access the current density information from a measured magnetic field), the Fourier transform of measured magnetic field can be divided by the Fourier transform of the Green's function, leaving the Fourier transform of the source current density[1]. Taking the inverse Fourier transform then returns the source current density components in real space. For finite size magnetic field sensors, one would also need to consider reversing the convolution of the magnetic field with that of the sensor function, however, NV-defects are atomic sized (approximating a Dirac delta function), so this convolution need not be considered.

In NV-magnetometry, only a single projection of the magnetic field is measured, meaning the full vector magnetic field information is not available. This projection lies along the axis connecting the nitrogen atom to the vacancy (termed the NV-quantisation axis), and can be any one of the <111> crystallographic axes of diamond. Its relation to the arbitrary lab axes $(x, y, z)$ depends on the cut of diamond from which the NV-tip is fashioned. The NV projection axis is represented in the cartesian frame using the spherical angles ($\theta$ is the azimuthal angle and $\phi$ the polar angle), by:

$$\boldsymbol{u} = \begin{pmatrix} u_x \\ u_y \\ u_z \end{pmatrix} = \begin{pmatrix} \sin\phi \cos\theta \\ \sin\phi \sin\theta \\ \cos\phi \end{pmatrix} \quad \text{(S2)}$$

This magnetic field information is sufficient to reconstruct both components of a 2D current density, $\boldsymbol{J} = (j_x, j_y, 0)$. The equations governing the reconstruction of a current density from a single projection of the magnetic field (measured in a plane above the source current) are given by Broadway et al.[2] as:

$$j_x = \frac{k_y}{g * (-u_x k_x - u_y k_y + i u_z k)} b_{\theta,\phi} \quad \text{(S3)}$$

and

$$j_y = \frac{k_x}{g * (u_x k_x + u_y k_y - i u_z k)} b_{\theta,\phi} \quad \text{(S4)}$$

Here, $j_x$ and $j_y$ are the Fourier transforms of the real-space current density components $J_x$ and $J_y$, $k_x$ and $k_y$ are the spatial frequencies along $x$ and $y$ (determined by the spatial extent of the magnetic field image), $k$ is the magnitude of the spatial frequency vector ($k = \sqrt{k_x^2 + k_y^2}$), $b_{\theta,\phi}$ is the Fourier transform of measured magnetic field, and $g$ is the Fourier transform of the Green's function relating current density and magnetic field:

$$g = \frac{\mu_0}{2} e^{-k z'} \quad (S5)$$

Here, $z'$ is the vertical distance between the source plane and the measurement plane. As can be seen from equations S2-S5, the current density is essentially a high-pass filtered version of the magnetic field. As such, errors in the reconstructed current can be induced due to[1,2]:

i) The loss of information associated with the effective filtering processes.
ii) The amplification of high frequency noise in the measured magnetic field data.
iii) Due the finite sampling area, which may exclude some magnetic field components that contain information about the source.

An example of the first error source is made clear when attempting to reconstruct a uniform current density in the $y$ direction from a measurement of the $y$ component of the magnetic field. In this case, the projection axis would be $\boldsymbol{u} = (0,1,0)$, and equation S4 becomes:

$$j_y = \frac{k_x}{g * u_y k_y} b_{\theta,\phi} \quad (S6)$$

A spatially uniform current in the y direction indicates that $k_y = 0$, so the filter diverges and the reconstruction is ill-defined. To combat these issues, the measured magnetic field axis should have both in-plane and out-of-plane components (in the lab reference frame). We use NV tips with azimuthal and polar angles of approximately $\theta = 80°$ and $\phi = 45°$, respectively, so this condition is fulfilled naturally. Secondly, to combat the amplification of high frequency noise, we apply a Hanning filter to the magnetic field data (in frequency space), as discussed in other works[1,2].

Finally, so called truncation artefacts can arise due to unseen magnetic field components, missed by the finite sampling window. These effects can be reduced by padding of the measured magnetic field data with zeros, with assumed exponential decay of fields at the edge of the imaged window, or with "replication padding", where values at the edge of the frame are repeated outward in all directions[2]. Here, we have used replication padding to generate square images which are 60% longer than the longest edge of the real data. To convert to a three-dimensional current density, the reconstructed current density maps were divided by the approximate thickness of the electrode (100nm), with uniformity along the thickness direction assumed.

All analysis was carried out in MATLAB, using in-built fast Fourier transform algorithms and Hanning window functions, along with a user written padding function.

## Section S3: Large scale PFM + NV Images of *in-situ* Device

Figure S2 shows PFM amplitude (**a**) phase (**b**) images, taken from the same device as those appearing in the NV-mapping and PFM mapping in figure 3 of the main text. Here, a larger view is given to show that conducting domain wall pathways do indeed extend over the entire electrode area. The approximate location of the current carrying platinum interconnect is marked by a gold rectangle. Larger scale NV-centre magnetometry (**c**) and corresponding current density (**d**) maps are shown, which show that no significant current density exists outside of the scan window covered in figure 3 of the main text. Note that there is a difference in scale and positioning of the PFM images compared to the NV mapping; the NV map is larger and covers more of the top electrode.

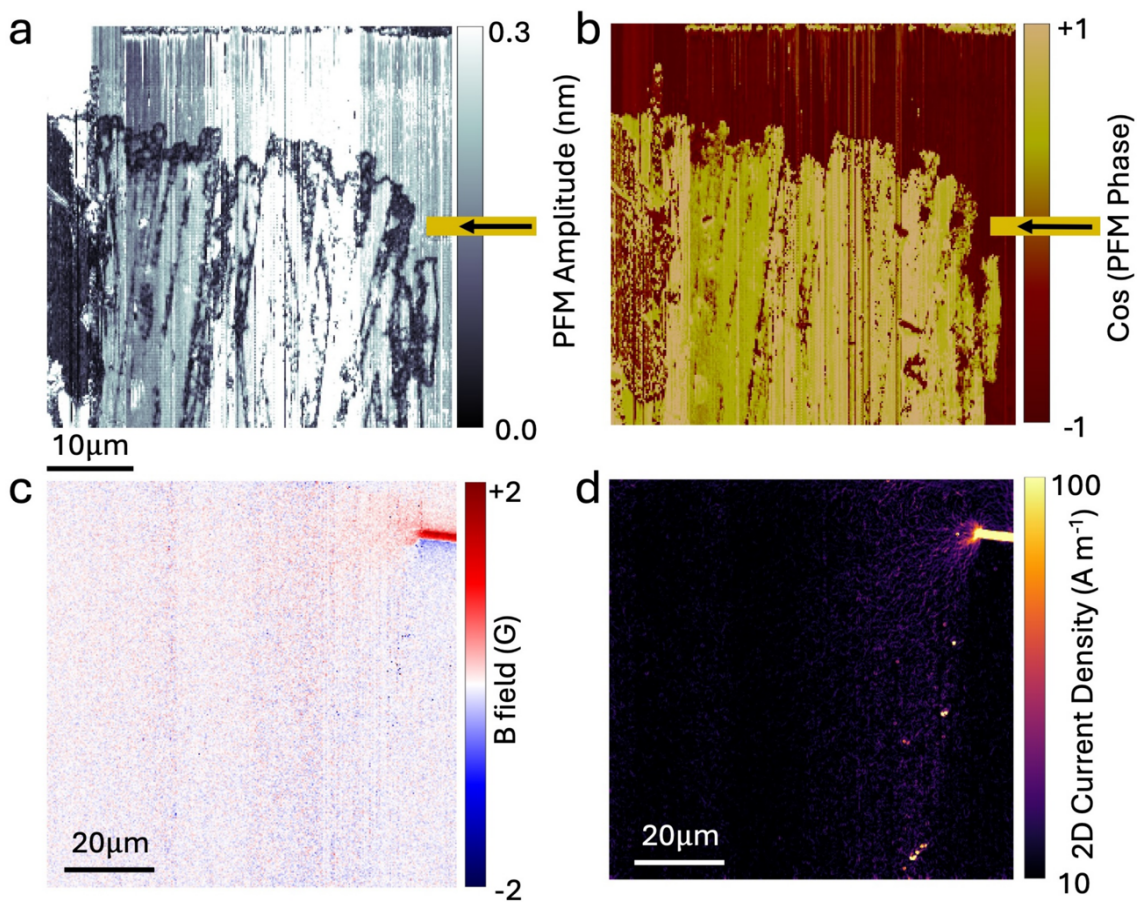

**Figure S2: Large scale domain mapping and NV-centre current mapping. a,** A PFM amplitude map (**a**), phase map (**b**) *in-situ* scanning NV-centre magnetic field map (**c**) reconstructed current density map (**d**) showing a larger view of the current carrying LNO capacitor structure.

## Section S4: Varying Conductivity in COMSOL Models

As discussed in the main text, the finite element modelling can replicate the measured current density profiles when only a small number of domain walls are active. We enforced this scenario in our finite element modelling by removing percolating current pathways other than a subset close to the current carrying platinum interconnect (figure 4a-c). The observations in figure 1 and figure S2 clearly show, however, that there are current pathways distributed over the entire top electrode area. Another way to replicate the observed current density pathways in the modelling is to increase the conductivity of the domain walls to match that of the metallic connections. In this case, current will mostly channel through the closest domain walls, as this offers the path of least resistance. This is illustrated in figure S3, where the model contains conducting domain wall pathways over the entire electrode area (as in figure 4e). Upon increasing the conductivity, current channelling similar to that observed in the experiment, could be recreated (figure **S3b**). In this case, however, the potential profile deviates from the equipotential profile experimentally observed in the KPFM scan in figure **2f**, as shown in figure **S3d** below.

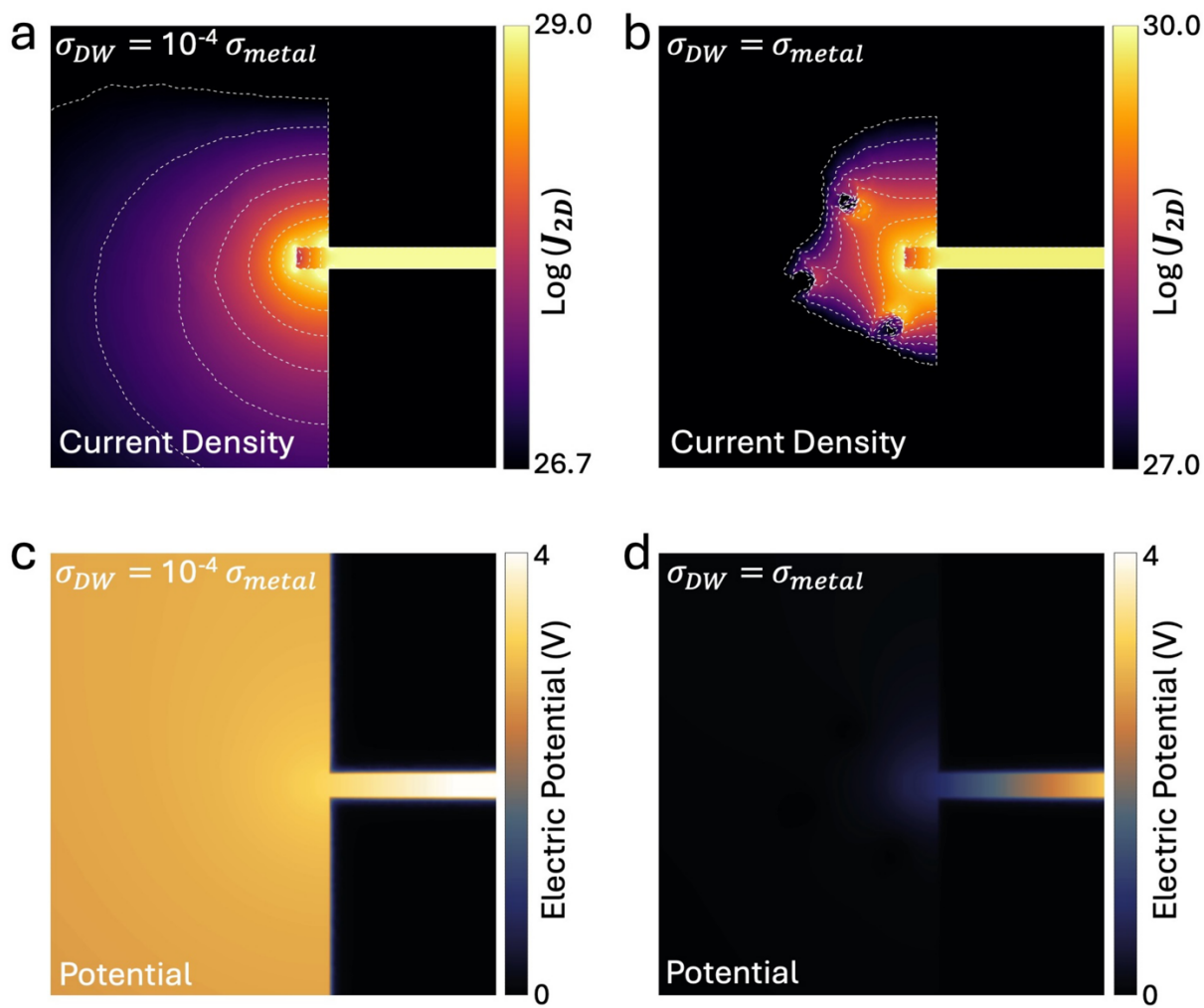

**Figure S3: Variable Domain Wall Conductivity in Finite Element Modelling. a, b,** Modelled Current density and **c, d,** potential profiles for simulated current flow through a domain wall capacitor with domain walls spread over the entire electrode area. In **a** and **c,** the domain wall conductivity is a factor of $10^4$ smaller than the metallic electrode, whereas in panels **b** and **d**, the domain wall conductivity matches that of the metallic electrodes.